\documentclass[10pt,twocolumn]{article} 
\usepackage{ol2}
\usepackage{hyperref}
\usepackage{amsmath}
\usepackage{amsfonts}
\usepackage{amsmath}
\usepackage{amssymb}
\begin{document}
\newcommand{\ud}{\mathrm{d}}
\newcommand{\ue}{\mathrm{e}}
\newcommand{\disp}{\displaystyle}
\newcommand{\der}{\partial}
\newcommand{\intl}{\int_{-\infty}^{+\infty}}
\newcommand{\imag}{\mathrm{Im}}
\newcommand{\sign}{\mathrm{sign}}
\twocolumn[
\title{\bf Dark solitons in mode-locked lasers}
\author{Mark J. Ablowitz$^1$, Theodoros P. Horikis$^2$, Sean D. Nixon$^1$,
Dimitri J. Frantzeskakis$^3$}
\address{$^1$Department of Applied Mathematics,
University of Colorado, 526 UCB, Boulder, CO 80309-0526\\
$^2$Department of Mathematics, University of Ioannina, Ioannina 45110, Greece\\
$^3$Department of Physics, University of Athens, Panepistimiopolis,
Zografos, Athens 15784, Greece}
\begin{abstract}
Dark soliton formation in mode-locked lasers is investigated by means of a
power-energy saturation model which incorporates gain and filtering saturated
with energy, and loss saturated with power. It is found that general initial
conditions evolve into dark solitons under appropriate requirements also met in
the experimental observations. The resulting pulses are well approximated by
dark solitons of the unperturbed nonlinear Schr\"{o}dinger equation. Notably,
the same framework also describes bright pulses in anomalous and normally
dispersive lasers.
\end{abstract}
\ocis{140.4050, 140.3510, 060.5530, 190.5530}
]   
\noindent Laser frequency stabilization via mode-locked (ML) lasers has become
an indispensable tool in many research activities. Advances in optical
frequency standards have resulted in the development of precise frequency
measurement capability in the visible and near-infrared spectral regions.
Although the potential for using ML lasers in optical frequency synthesis was
recognized early, the available lasers did not provide the properties necessary
for fulfilling this potential until recently \cite{cundiff2}. The recent
explosion of measurements based on ML lasers has been largely due to the
development of the Kerr-lens mode-locked ti:sapphire laser and its capability
to generate sufficiently short pulses. Alternatively, amplified pulses of a
similar duration are created through the compression of pulses via self-phase
modulation in a hollow core fiber \cite{zhang}.

Dark solitons are intensity dips on a constant background with a phase jump
across their intensity minimum. Since their discovery, cf. \cite{dark1973},
dark solitons have attracted considerable attention, especially in the fields
of nonlinear optics \cite{kivpr} and Bose-Einstein condensates \cite{dfrantz}.
In the non-mode-locking regime, the first train of dark solitons was
successfully achieved in an all normal dispersion erbium-doped fiber laser with
an in-cavity polarizer. In ML lasers they have been difficult to generate with
the first experimental observation only recently reported \cite{zhang}; see
also \cite{cundiff_preprint}. Apart from the bright pulse emission, it was
observed that, under under strong continuous wave emission, appropriate pump
strength and negative cavity feedback, a fiber laser can also emit single or
multiple dark pulses \cite{zhang}.

To analyze dark solitons in ML lasers we use a model based on a variant of the
nonlinear Schr\"odinger (NLS) equation, termed the power-energy saturation
(PES) equation \cite{horikis}, suitably generalized to include non-vanishing
boundary conditions at infinity. It is expressed in the following dimensionless
form,
\begin{gather}
i \psi_z - \frac{1}{2} \psi_{tt} +|\psi|^2\psi =\nonumber\\
\frac{ig}{1+E/E_0}\psi + \frac{i\tau}{1+E/E_0}\psi_{tt} -
\frac{il}{1+P/P_0}\psi,
\label{pes}
\end{gather}
where the complex electric field envelope $\psi(z,t)$ is subject to the
boundary conditions $|\psi(z,t)| \rightarrow |\psi_{\infty}|$ as $|t|
\rightarrow \infty$. Here, $E(z)=\intl (|\psi_\infty|^2-|\psi|^2)\;\ud t$ is
the dark-pulse energy, $P(z,t)=|\psi_\infty|^2-|\psi|^2$ is the instantaneous
power, while $E_0$ and $P_0$  are related to the saturation energy and power,
respectively. Furthermore, $g$, $\tau$, and $l$ are all positive, real
constants, with the corresponding terms representing saturable gain, spectral
filtering, and saturable loss. Typical dimensional numbers can be found in Ref.
\cite{horikis3}.

The above definition of the pulse energy, although unconventional, corresponds
to the physical properties of the system. Indeed, as the peak-to-background
intensity ratio can achieve large values in bright pulses, so can the
focused-to-unfocused vacuum ratio in dark pulses. As such, dark pulses can be
thought of as focusing the vacuum \cite{volpe}. The definition of power follows
consistently from $E=\intl P\;\ud t$. A modified model (yielding similar
results), with more standard definitions of energy and power, is also discussed
below.

\begin{figure}[!htbp]
    \centering
    \includegraphics[height=2in]{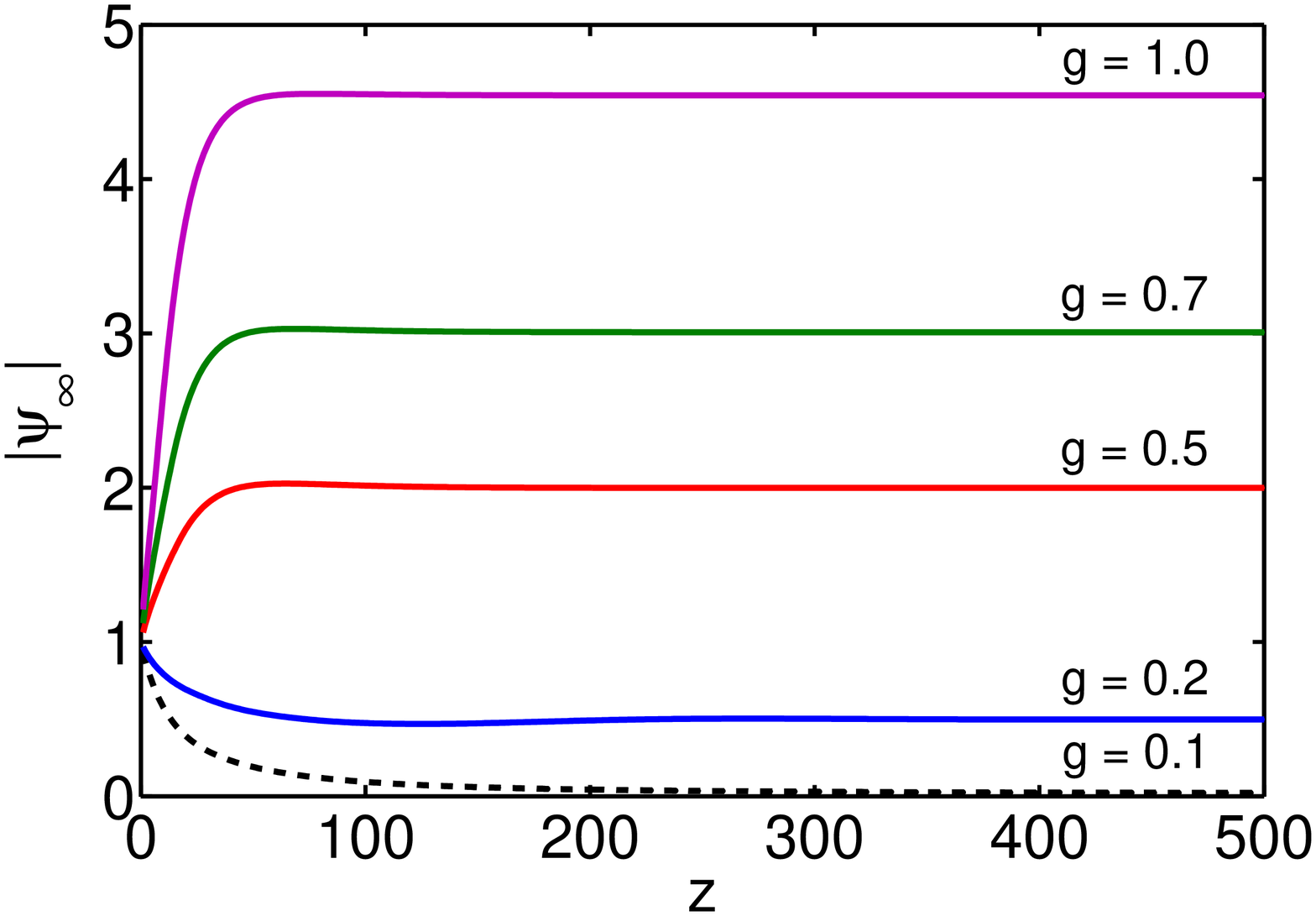}
    \includegraphics[height=2in]{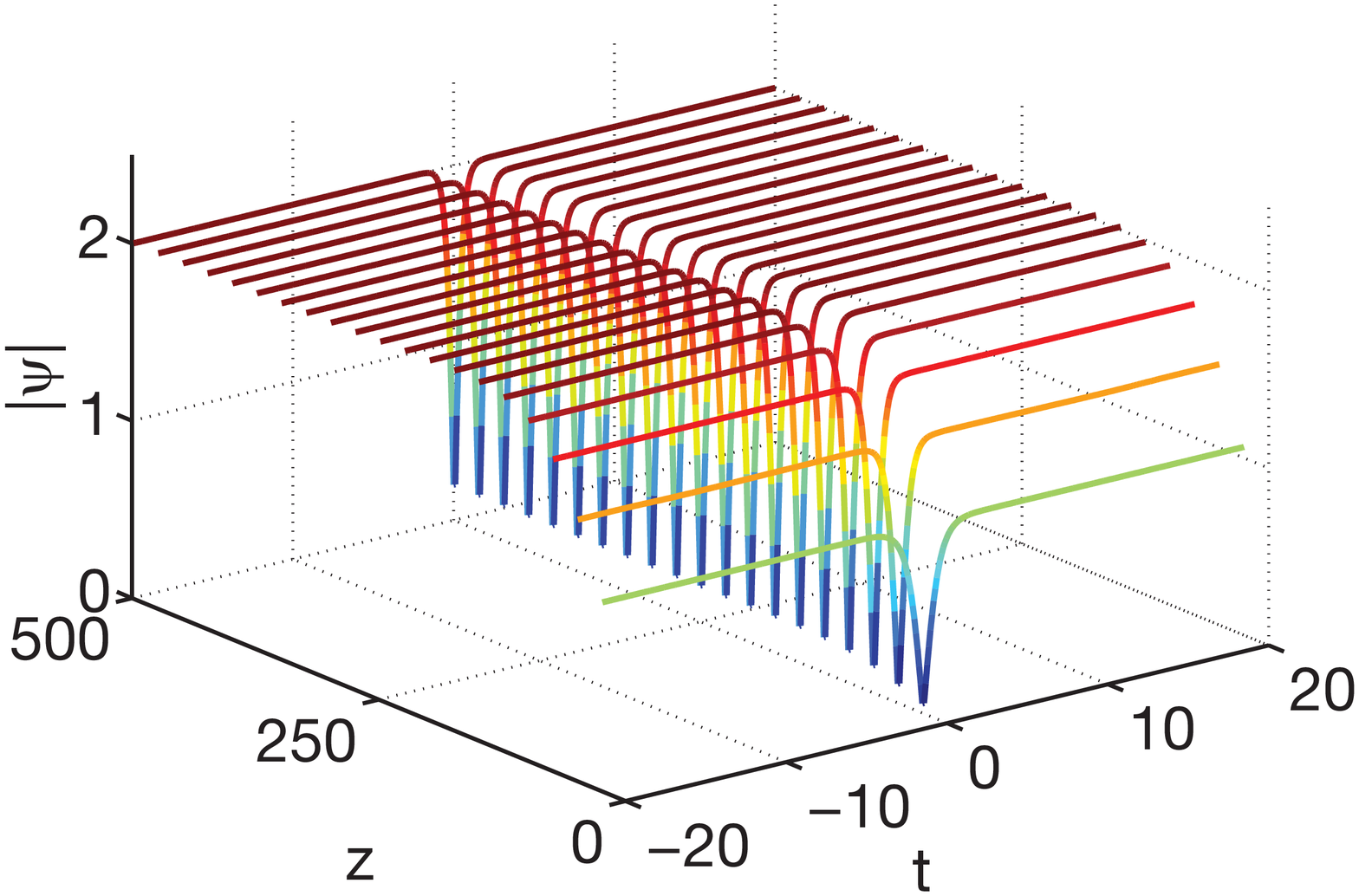}
    \caption{(Color online) Top panel: evolution of the pulse background of an
    arbitrary initial profile under the PES equation with different values of gain.
    Bottom panel: complete evolution of the initial profile for $g=0.5$.}
    \label{evol_arb}
\end{figure}

To study the dynamics and mode-locking capabilities of our model, we integrate
Eq. (\ref{pes}) with a fourth-order Runge-Kutta method, with initial profile
$\psi(0,t)=\tanh t$, though others, with a $\pi$-phase jump, can be used as
well. The gain parameter $g$ is varied, while $E_0=P_0=1$ and $\tau=l=0.1$. In
the top panel of Fig. \ref{evol_arb} we show the evolution of the background
amplitude, which sets the dark soliton amplitude, for different values of the
gain parameter $g$.
Locking onto stable dark solitons is only achieved when the gain term is
sufficiently strong, i.e. the parameter $g$  is large enough to counter balance
the losses. Thus, for $g=0.1$ (see dashed line in the top panel of
Fig.~\ref{evol_arb}), the dark soliton decays quickly due to excessive loss. On
the other hand, for $g=0.2$, the pulse initially decreases, but then locks to a
specific amplitude and width. Finally, for $g=0.5$, $0.7$, $1.0$, a stable
evolution characterized by an increase of the soliton amplitude is obtained
almost instantaneously; see bottom panel of Fig.~\ref{evol_arb}.

To determine the evolution of the background wave in the framework of
Eq.~(\ref{pes}), we assume that $\psi(z)=\psi_\infty(z)\exp[i\theta(z)]$.
Separating real and imaginary parts, yields the following equation for the
background amplitude $\psi_\infty(z)$:
\begin{equation}
\frac{\ud\psi_\infty}{\ud
z}=\frac{g}{1+2|\psi_\infty|/E_0}\psi_\infty-l\psi_\infty. \label{psi_infty}
\end{equation}
Here, an approximate solution of Eq.~(\ref{pes}) in the form $|\psi(z,t)| =
|\psi_\infty\tanh(\psi_\infty t)|$ is assumed, which gives $E=2|\psi_\infty|$.
Equation \eqref{psi_infty} also describes the evolution of $\psi_\infty$, as
depicted in the top panel of Fig. \ref{evol_arb}, and illustrates the
importance of energy saturation. Indeed, if we consider the so-called
similariton supporting equation (NLS with linear gain) corresponding to
$\tau=l=0$ \cite{boscolo,horikis3}, with energy saturation absent
($2|\psi_\infty|/E_0=0$), the pulse amplitude grows exponentially at a rate
defined by the gain $g$. Contrary, when saturation is present, this
exponentially growing state cannot be reached. If fact, all terms in Eq.
\eqref{pes} are important for successful mode-locking. Spectral filtering is
important in maintaining a short pulse duration, which is a generic phenomenon
to mode-locking in the normal dispersion regime \cite{bale2}. With spectral
filtering in the cavity, stable ultrashort pulses can be generated at very high
values of the group-velocity dispersion. Saturable loss results in higher net
gain for a train of short pulses (compared to continuous-wave operation), which
is also necessary for mode-locking. In some models, loss is introduced in the
form of fast saturable power absorbers placed periodically \cite{ilday1}. It
has been shown that the lumped and distributive models yield similar results
\cite{horikis_epj}.

\begin{figure}[!htbp]
    \centering
    \includegraphics[height=2in]{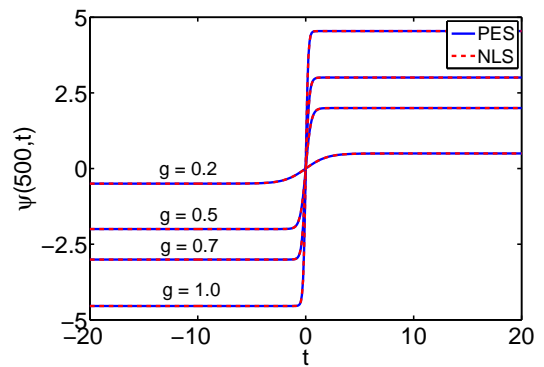}
    \caption{(Color online) Black solitons of the PES and NLS
    equations for different values of $g$.}
    \label{nls.compare}
\end{figure}

A stable equilibrium (attractor) exists for Eq. \eqref{psi_infty} and can be
found setting $\ud \psi_\infty / \ud z =0$, namely,
\begin{gather}
|\psi_\infty|=\frac{E_0}{2}\left( \frac{g}{l}-1 \right). \label{energy}
\end{gather}
Equation (\ref{energy}) is the resulting background amplitude of the dark
soliton and agrees with direct numerical simulation, as seen in Fig.
\ref{nls.compare}. Thus, dark solitons tend to an equilibrium (mode-lock) with
constant energy and background.

The above findings agree with the experiment of Ref.~\cite{zhang}, where the
mode-locked dark pulses were identified as NLS dark solitons. In particular,
they are {\em black} solitons (stationary kinks) characterized by a $\pi$-phase
jump across the soliton notch and they can be described analytically by a
hyperbolic tangent profile, i.e., $\psi(z,t)=\psi_\infty \tanh(\psi_\infty t)$.
In Fig. \ref{nls.compare} we plot the black solitons of the PES and NLS
equations (for the same value of $\psi_\infty$, as determined from Eq.
\eqref{energy}) for different values of $g$.
The amplitudes match so closely that they are indistinguishable in the figure,
meaning that the combined perturbation of gain, filtering and loss in the PES
model underlies the mode-locking mechanism: its effect is to mode-lock to a
black soliton of the pure NLS equation with the appropriate background,
Eq.~\eqref{energy}. On the other hand, mode-locking to {\em grey} solitons,
namely moving dark pulses with a phase-jump less than $\pi$ at their (non-zero)
intensity minima, are not found in this model as they have insufficient
dark-pulse energy. This too agrees with Ref.~\cite{zhang} where grey solitons
were not observed. Note that recent experimental \cite{cundiff} and theoretical
\cite{horikis2} results in ML lasers in the anomalous dispersion regime
indicate that the normalized intensity of a pulse in a ML laser can be
described by the bright solitons of the unperturbed NLS equation.

\begin{figure}[!htbp]
    \centering
    \includegraphics[height=2in]{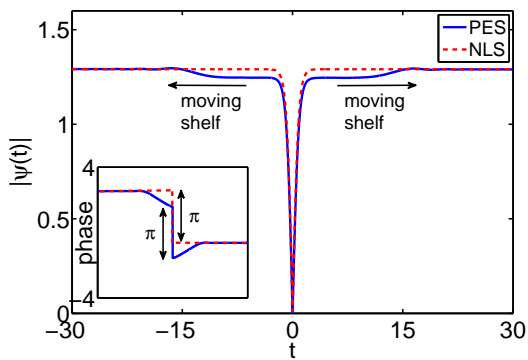}
    \caption{(Color online) The development of a shelf in the PES equation
    (solid line) as compared to the solution of the NLS equation (dashed line).}
    \label{nls.shelf}
\end{figure}

The conventional definition of the pulse energy may also be used in Eq.
\eqref{pes}, namely $E=\frac{1}{T}\int_{-T/2}^{T/2} |\psi|^2\;\ud
t$, where $T$ is the averaged time
through the cavity. In this case, looking for stationary solutions of
Eq.~\eqref{energy} yields
\begin{gather}
|\psi_\infty|^2=\frac{g-l}{l/E_0-g/P_0} \label{background.new}
\end{gather}
for the background where, for sufficiently large $T$,
$E=\frac{1}{T}\int_{-T/2}^{T/2} |\psi|^2\;\ud t \approx
\frac{1}{T}|\psi_\infty|^2 \,T=|\psi_\infty|^2$. Equation
\eqref{background.new} is a more restrictive condition for mode-locking which
can also be attributed to the experimental difficulties of the problem.
Furthermore, this definition results in more complicated dynamics and the
development of a {\it shelf} on the soliton background, which is more
pronounced as compared to the previous case. In fact, as shown elsewhere
\cite{nixon}, shelves occur naturally in the perturbation theory for dark
solitons. We illustrate the resulting shelf in Fig.~\ref{nls.shelf}, where now
$g=0.2$, $l=0.1$, $E_0=1$ and $P_0=5$.
Due to the noisy background this feature maybe difficult to observe in an
experiment. However, the shelf also affects the phase of the resulting pulse
(see inset in Fig. \ref{nls.shelf}) and since the experimentally observed
pulses exhibit a sharp $\pi$-phase jump \cite{zhang}, the first dark-pulse
energy definition may give a closer approximation for this problem.

\begin{figure}[!htbp]
    \centering
    \includegraphics[height=2in]{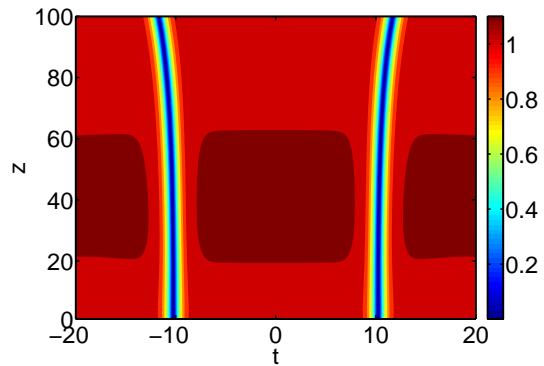}
    \caption{(Color online) Two-dark-soliton evolution.}
    \label{two.shelf}
\end{figure}

Lastly, we mention the notion of multiple dark solitons. To observe these, we
evolve Eq.~\eqref{pes} with initial condition
$\psi(0,t)=\tilde\psi_\infty\tanh(\tilde\psi_\infty t+t_0)$ for $t<0$ and
$\psi(0,t)=-\tilde\psi_\infty\tanh(\tilde\psi_\infty t-t_0)$ for $t>0$, where,
$\tilde\psi_\infty$ can be approximated by $\psi_\infty/2$; the dark-pulse
energy is now twice that of the single soliton. In Fig.~\ref{two.shelf} we show
its evolution.
As in the case of bright pulses \cite{nixon2}, the two solitons are taken
initially sufficiently far apart (about 10 times their full-width of
half-maximum) to minimize interactions. The major difference with the bright
case, where the interaction is logarithmically slow \cite{nixon2}, is that
pulses repel sooner here due to the shelf interactions. While the solitons are
stable for reasonably long distance enough to also be experimentally observed,
nevertheless one can expect this case to be sensitive as with the single dark
soliton.

In conclusion, we presented a power-energy saturation model that describes
mode-locking of dark solitons in lasers. Much like their bright counterparts,
these pulses (as well as multiple pulses) exist when sufficient gain is present
in the system, which agrees with the experimental observations. The specific
energy saturated gain and filtering, and power saturated loss, are crucial to
the mode-locking mechanism. The resulting pulses are essentially modes of the
unperturbed NLS equation with background amplitudes appropriately defined by
the gain and loss parameters.

This research was partially supported by the U.S. Air Force Office of
Scientific Research, under grant FA9550-09-1-0250.

\bibliographystyle{osajnl}

\end{document}